\documentclass[aps,pra,nofootinbib,twocolumn]{revtex4-2}
\usepackage{amsmath, amsthm, amscd, amssymb}
\usepackage{bm}
\usepackage{bbm}
\usepackage{epsfig}

\begin{document}

\title{Nonthermal radiation of evaporating black holes}
\author{Anatoly A. Svidzinsky }
\affiliation{Texas A\&M University, College Station, TX 77843}

\begin{abstract}
Black hole (BH) evaporation is caused by creation of entangled
particle-antiparticle pairs near the event horizon, with one carrying
positive energy to infinity and the other carrying negative energy into the
BH. Since under the event horizon, particles always move toward the BH
center, they can only be absorbed but not emitted at the center. This breaks
absorption-emission symmetry and, as a result, annihilation of the particle
at the BH center is described by a non-Hermitian Hamiltonian. We show that
due to entanglement between photons moving inside and outside the event
horizon, non-unitary absorption of the negative energy photons near the BH
center, alters the outgoing radiation. As a result, radiation of the
evaporating BH is not thermal, it carries information about BH interior and
entropy is preserved during evaporation.
\end{abstract}

\date{\today }
\maketitle

\section{Introduction}

According to principles of quantum mechanics, state of an isolated system
remains pure during evolution. This is the case for both types of quantum
mechanical evolution - a unitary evolution governed by the Schr\"{o}dinger
equation and a non-unitary state vector collapse brought about by a
measurement. If the system remains in a pure state the von Neumann entropy
is preserved.

Computations of Hawking radiation, which is believed to be produced by an
evaporating black hole (BH), indicated that it is completely thermal \cite%
{Hawk74,Hawk75,Mukh07,Frol98}. Therefore, an evaporating BH would eventually
leave behind a cloud of thermal radiation, independently of the initial
state from which it was formed. However, one could imagine forming a BH from
a pure state that seems to evolve to a mixed thermal state which amounts to
a loss of information and thus is incompatible with quantum mechanical
evolution. This is known as the BH information paradox \cite{Hawk76}. For
proposals to resolve the BH information problem see, e.g., \cite%
{Maro17,Calm22b} and references therein.

According to the holographic principle, the bulk information in models of
gravity in $d$-dimensions might be available on the $d-1$ dimensional
boundary of spacetime \cite{Hoof93,Suss95}. Holography of information
implies that the internal quantum state of a BH must be encoded in the
asymptotic quantum state of its graviton field, since otherwise the
information would not be recoverable at the boundary \cite{Ladd21,Chow21}.
In \cite{Clam22a,Clam22} it has been shown explicitly that information about
the BH internal state is available in the quantum state of its gravity field
(quantum hair). Moreover, it has been argued that long wavelength gravitons
can give rise to an infinite number of conserved charges which preserve an
infinite amount of information outside BHs \cite{Stro14} which could give a
new perspective on the information problem \cite{Hawk16,Hawk17}.

Holography was given an explicit realization in the AdS/CFT correspondence
of Maldacena \cite{Malda98}, which suggests that BH evaporation can be
unitary. Recently there has been considerable progress in directly computing
the entanglement entropy of evaporation using AdS methods, and these results
suggest that the process is unitary \cite{Almh21}. While both holography of
information and AdS/CFT duality suggest that the BH information paradox is
somehow resolved in favor of unitarity, neither yield a specific description
of the physical process by which BH information is encoded in Hawking
radiation.

Here we show that entanglement of particle pairs generated during BH
evaporation, combined with non-unitary absorption of particles near the BH
center, leads to nonthermal outgoing radiation that carries information
about the BH interior.

BHs possess an event horizon - the boundary under which no particles, at
least if they are treated classically and moving forward in time, can
escape. This leads to a believe that an observer outside the BH has no
access to the interior part of the total quantum system, and information
about the internal degrees of freedom is lost during BH evaporation leaving
the system in a mixed state.

However, BH spacetime has another inherent feature, namely, under the event
horizon, particles always move toward the BH center. The latter probably has
a Planck scale and is described by yet not well-understood physics. What
matters for the present discussion is that, since particles can move only
towards the center, photons with a wavelength much greater than the Plank
length can only be absorbed, but not emitted at the BH center.

Emitted particles move away from the source and, since particles cannot move
away from the BH center, they cannot be emitted in this region. This breaks
the symmetry between absorption and emission. As a result, annihilation of
the particle at the BH center is described by a non-Hermitian Hamiltonian.

Here we show that if the emission-absorption symmetry is broken, quantum
mechanics predicts that radiation of an evaporating BH is nonthermal and it
carries information about the state of matter in the BH interior. Next we
briefly discuss the physics of Hawking radiation from a negative frequency
perspective \cite{Svid21}.

\begin{figure}[h]
\begin{center}
\epsfig{figure=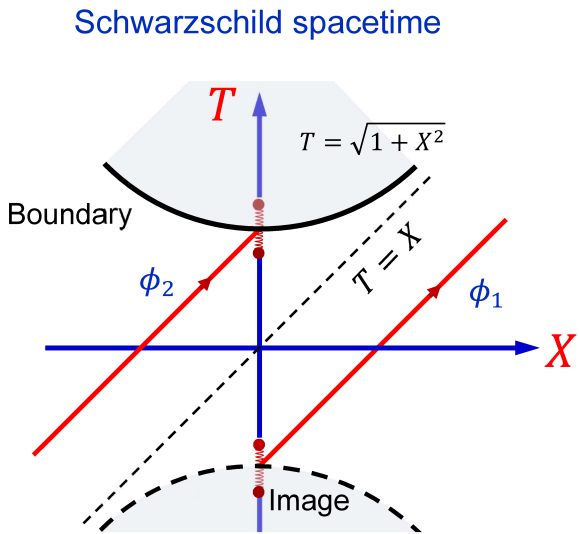, angle=0, width=9cm}
\end{center}
\caption{Schwarzschild spacetime in the Kruskal-Szekeres coordinates. A
space-like line $T=\protect\sqrt{1+X^{2}}$ sets the spacetime boundary.
Unruh vacuum is filled with entangled right-moving Rindler photons $\protect%
\phi _{1}$ and $\protect\phi _{2}$ which are localized outside and inside
the BH event horizon (line $T=X$) respectively. Absorption of Rindler
photons $\protect\phi _{2}$ at the boundary reduces BH mass and leads to BH
evaporation. We model the boundary as a set of harmonic oscillators that
absorb all ingoing photons, but do not emit. Due to vacuum entanglement, the
process looks like as if there is a mirror \textquotedblleft
image\textquotedblright\ of the oscillators located along the line $T=-%
\protect\sqrt{1+X^{2}}$ which emit (but do not absorb) light outside the BH
event horizon. }
\label{Fig1}
\end{figure}

\section{Hawking radiation from a negative frequency perspective}

According to general relativity, a static BH of mass $M$ in 3+1 dimension in
Schwarzschild coordinates is described by a metric%
\begin{equation}
ds^{2}=\left( 1-\frac{r_{g}}{r}\right) c^{2}dt^{2}-\frac{1}{1-\frac{r_{g}}{r}%
}dr^{2}-r^{2}\left( d\theta ^{2}+\sin ^{2}\theta d\varphi ^{2}\right) ,
\label{q4}
\end{equation}%
where $r_{g}=2GM/c^{2}$ is the gravitational radius. For simplicity, we
truncate the spacetime to 1+1 dimension ($t$ and $r$, where $r\geq 0$) and
use Kruskal-Szekeres coordinates $T$ and $X$ that are defined in terms of
the Schwarzschild coordinates $t$ and $r$ as%
\begin{equation}
T=\sqrt{r/r_{g}-1}e^{\frac{r}{2r_{g}}}\sinh \left( \frac{ct}{2r_{g}}\right) ,
\label{k1}
\end{equation}%
\begin{equation}
X=\sqrt{r/r_{g}-1}e^{\frac{r}{2r_{g}}}\cosh \left( \frac{ct}{2r_{g}}\right) ,
\label{k2}
\end{equation}%
for $r>r_{g}$, and%
\begin{equation}
T=\sqrt{1-r/r_{g}}e^{\frac{r}{2r_{g}}}\cosh \left( \frac{ct}{2r_{g}}\right) ,
\label{k3}
\end{equation}%
\begin{equation}
X=\sqrt{1-r/r_{g}}e^{\frac{r}{2r_{g}}}\sinh \left( \frac{ct}{2r_{g}}\right) ,
\label{k4}
\end{equation}%
for $0<r<r_{g}$. In these coordinates, the BH center ($r=0$) is a space-like
line $T=\sqrt{1+X^{2}}$. This line sets a boundary of the Schwarzschild
spacetime in the Kruskal-Szekeres coordinates (see Fig. \ref{Fig1}). The
boundary appears because coordinate transformation (\ref{k1})-(\ref{k4})
maps the region $-\infty <t<\infty $, $r\geq 0$ into $T\leq \sqrt{1+X^{2}}$, 
$T\geq -X$. In the Kruskal-Szekeres coordinates, in 1+1 dimension, the
Schwarzschild metric%
\begin{equation}
ds^{2}=\frac{4r_{g}^{3}}{r}e^{-r/r_{g}}\left( dT^{2}-dX^{2}\right) 
\label{q5}
\end{equation}%
is conformally invariant to the Minkowski metric and, thus, a massless
scalar field $\phi $ obeys the same wave equation as in the Minkowski
spacetime%
\begin{equation}
\left( \frac{\partial ^{2}}{\partial T^{2}}-\frac{\partial ^{2}}{\partial
X^{2}}\right) \phi =0.  \label{q8}
\end{equation}

For the present problem only the right-moving field in Fig. \ref{Fig1} is
important. It is convenient to describe such a field using Rindler modes 
\cite{Rind66} 
\begin{equation}
\phi _{1\Omega }(T,X)=(X-T)^{i\Omega }\theta (X-T),  \label{R1a}
\end{equation}%
\begin{equation}
\phi _{2\Omega }(T,X)=(T-X)^{-i\Omega }\theta (T-X),  \label{R1b}
\end{equation}%
where $\Omega >0$ is a parameter, and $\theta $ is the Heaviside step
function. Rindler modes $\phi _{1\Omega }$ and $\phi _{2\Omega }$ are
solutions of the wave equation (\ref{q8}), and for $\Omega >0$ have positive
norm (defined as the Klein--Gordon inner product). The mode functions (\ref%
{R1a}) and (\ref{R1b}) are non-zero outside and inside the BH event horizon
(line $T=X$) respectively (see Fig. \ref{Fig1}). These two regions are
causally disconnected for the right-moving field. Annihilation operators of
the Rindler photons we denote as $\hat{b}_{1\Omega }$ and $\hat{b}_{2\Omega }
$.

It is believed that, to a good approximation, Unruh vacuum $|0_{U}\rangle $
describes state of the field produced by a gravitational collapse of a star
into a BH. In this state, there are no left-moving Rindler photons and no
right-moving Minkowski photons \cite{Mukh07}. That is, Unruh vacuum is
Rindler vacuum for the left-moving photons and Minkowski vacuum for the
right-moving photons. In terms of the right-moving Rindler photons, which
are relevant for the present discussion, the Unruh vacuum is a squeezed
state \cite{Unru84} 
\begin{equation}
\left\vert 0_{U}\right\rangle =\prod\limits_{\Omega >0}\sqrt{1-\gamma ^{2}}%
e^{\gamma \hat{b}_{1\Omega }^{\dagger }\hat{b}_{2\Omega }^{\dagger
}}\left\vert 0_{R}\right\rangle ,  \label{RE}
\end{equation}%
where%
\begin{equation}
\gamma =e^{-\pi \Omega },  \label{k5}
\end{equation}%
$\left\vert 0_{R}\right\rangle $ refers to the Rindler vacuum, $\hat{b}%
_{1\Omega }^{\dagger }$ and $\hat{b}_{2\Omega }^{\dagger }$ are creation
operators of the right-moving Rindler photons. That is, Unruh vacuum is
filled with the right-moving Rindler photons, but it looks empty if the
right-moving field is described by means of the Minkowski photons.

If a hypothetical observer is located at the BH center ($r=0$), then
Schwarzschild coordinate $t$ is the proper time for such observer. Recall
that proper time of an object is the coordinate which changes in the
object's frame. If the object is held fixed at $r=const$ then $t$ is the
proper time. In the region $0<r<r_{g}$, it is physically impossible to hold
particles fixed at $r=const$, that's why we use the word \textquotedblleft
hypothetical\textquotedblright . Eqs. (\ref{k3}), (\ref{k4}) and (\ref{R1b})
yield that at the BH center the non-zero Rindler mode $\phi _{2\Omega }$
oscillates as a function of $t$ as $\phi _{2\Omega }\propto e^{i\Omega
ct/2r_{g}}$. That is, from the observer's perspective the Rindler photons
behave as if they have negative frequency $-\Omega c/2r_{g}$ \cite{Svid21}.
Hence, in the Unruh vacuum, there is a flux of the negative frequency
(energy) Rindler photons into the BH center. Absorption of such photons near
the BH center decreases energy (mass) of the BH, leading to BH evaporation.

If an observer is held fixed outside the event horizon at a constant
Schwarzschild coordinate $r$, then at the observer's location the non-zero
Rindler modes $\phi _{1\Omega }$ oscillate as $\phi _{1\Omega }\propto
e^{-i\Omega ct/2r_{g}}$, where $t$ is the observer's proper time. That is,
from the external observer perspective the Rindler photons behave as if they
have positive frequency%
\begin{equation}
\nu =\frac{\Omega c}{2r_{g}}  \label{k6}
\end{equation}%
and, thus, they can excite a detector. Photons $\phi _{1\Omega }$ propagate
away from the BH.

For simplicity, we will assume that the field has only modes with one
\textquotedblleft frequency\textquotedblright\ $\Omega $. We denote such
Rindler modes as $\phi _{1}$ and $\phi _{2}$. Then Unruh vacuum can be
written as 
\begin{equation}
\left\vert 0_{U}\right\rangle =\sqrt{1-\gamma ^{2}}e^{\gamma \hat{b}%
_{1}^{\dagger }\hat{b}_{2}^{\dagger }}\left\vert 0_{R}\right\rangle =\sqrt{%
1-\gamma ^{2}}\sum_{n=0}^{\infty }\gamma ^{n}\left\vert nn\right\rangle ,
\label{b1}
\end{equation}%
where $\left\vert nn\right\rangle $ is a state with $n$ Rindler photons in
the modes $\phi _{1}$ and $\phi _{2}$. 

If modes $\phi _{1}$ and $\phi _{2}$ are considered separately, then tracing
over or absorbing one of the modes leaves the remaining mode in a thermal
state. Namely, if we trace over the Rindler modes under the event horizon $%
\phi _{2}$, which are not accessible to the external observer, the reduced
density operator for the field $\phi _{1}$ is thermal 
\begin{equation}
\hat{\rho}_{1}=\text{Tr}_{2}\left( \left\vert 0_{U}\right\rangle
\left\langle 0_{U}\right\vert \right) =\left( 1-\gamma ^{2}\right)
\sum_{n=0}^{\infty }\gamma ^{2n}\left\vert n\right\rangle \left\langle
n\right\vert  \label{bT}
\end{equation}%
with the average number of photons%
\begin{equation}
\bar{n}_{1}=\frac{\gamma ^{2}}{1-\gamma ^{2}}.  \label{k7}
\end{equation}%
Thus, an observer held fixed outside the BH horizon feels thermal radiation
coming out from the BH, which is known as Hawking radiation. Using\ Eqs. (%
\ref{k5}), (\ref{k6}) and (\ref{k7}), one can write $\bar{n}_{1}$ as a
Planck factor 
\begin{equation}
\bar{n}_{1}=\frac{1}{e^{\frac{4\pi r_{g}\nu }{c}}-1}=\frac{1}{e^{\frac{\hbar
\nu }{k_{B}T_{H}}}-1}  \label{k8}
\end{equation}%
with the Hawking temperature $T_{H}=\hbar c/4\pi k_{B}r_{g}$.

\begin{figure}[h]
\begin{center}
\epsfig{figure=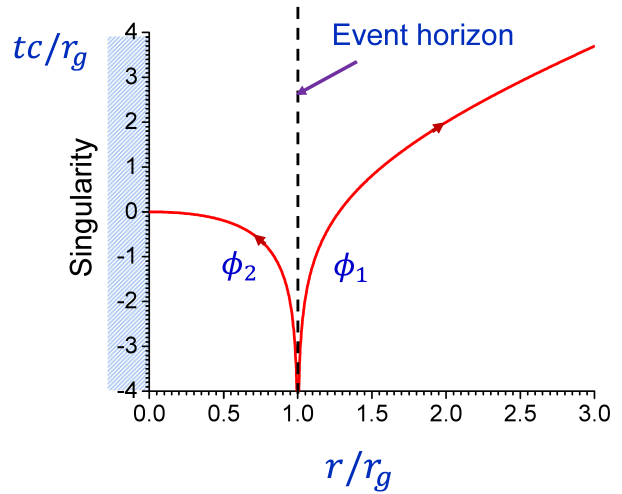, angle=0, width=9cm}
\end{center}
\caption{Light rays of Rindler photons $\protect\phi _{1}$ and $\protect\phi %
_{2}$ in Schwarzschild coordinates. }
\label{Fig2}
\end{figure}

Figure \ref{Fig2} shows light rays of Rindler photons (\ref{R1a}) and (\ref%
{R1b}) in the Schwarzschild coordinates. It looks like the negative ($\phi
_{2}$) and positive ($\phi _{1}$) frequency Rindler photons are generated at
the event horizon. This is consistent with the interpretation of the Hawking
radiation as a continuous creation of particle-antiparticle pairs near the
event horizon, with one carrying positive energy to infinity and the other
carrying negative energy into the BH \cite{Hawk77}. Calculations of the
energy-momentum tensor for the field near an evaporating BH directly show
that there is a negative-energy flux into the BH center and a
positive-energy flux far away from the BH \cite{Davi76}.

\section{Model of an evaporating black hole taking into account non-unitary
photon absorption at the center}

According to Eq. (\ref{q8}), in the Kruskal-Szekeres coordinates the field
evolves following the same wave equation as in Minkowski spacetime. For the
latter, absorption or emission of photons in the region $T>X$ cannot affect
the state of the field in the region $T<X$. However, the BH spacetime has a
space-like boundary at $T=\sqrt{1+X^{2}}$. We will show below that
absorption of photons at the boundary changes the state of the field outside
the event horizon and radiation of the evaporating BH is not thermal. We
will assume that Unruh vacuum is the state of the field only at the onset of
evaporation and calculate how the field evolves.

According to general relativity, spacetime disappears at the BH center
(spacetime boundary). It is assumed that matter disappears together with the
spacetime, but state of matter (mass, angular momentum, etc.) is recorded in
the gravitational field near the BH center. This process transfers
characteristics of the accreting matter into the BH internal gravitational
field.

The worldlines of the Rindler photons $\hat{b}_{2}$ terminate at the
spacetime boundary (see Fig. \ref{Fig1}). But we can't just say that photons
disappear. One should describe this process quantum mechanically using a
Hamiltonian. Space-like boundary breaks the symmetry between emission and
absorption of Rindler photons $\hat{b}_{2}$. Namely, if backward in time
propagation is not allowed, Rindler photons $\hat{b}_{2}$ cannot be emitted
at the boundary because such a process means emission of particles outside
the spacetime.

Next we consider a simple toy model of BH evaporation modeling the boundary
as a set of harmonic oscillators that totally absorb the ingoing field. The
oscillators follow the worldline of the boundary which is not geodesic. We
do not associate the oscillators with ordinary particles. Rather, the
oscillators provide a physical model of the gravitational field near the BH
center that carries information about the state of the BH interior. In our
model, the oscillator's energy is the origin of the BH mass. As we showed
above, from the oscillator's perspective, Rindler photons have negative
energy. Thus, absorption of Rindler photons reduces the energy of the
oscillators (BH mass decreases).

Since oscillators are under the BH horizon, they can interact only with
photons $\hat{b}_{2}$. In the toy model, the interaction Hamiltonian
describing BH evaporation reads%
\begin{equation}
\hat{V}_{2}(t)=g\hat{\sigma}e^{-i\omega t}\phi _{2}(t)\hat{b}_{2},
\label{b2a}
\end{equation}%
where $\hat{\sigma}$ is the lowering operator for the oscillator of
frequency $\omega $, $g$ is the coupling constant and the field mode
function $\phi _{2}(T,X)$ is taken at the location of the oscillator $\phi
_{2}(t)=\phi _{2}(T(t,0),X(t,0))=e^{i\frac{c\Omega t}{2r_{g}}}$. In Eq. (\ref%
{b2a}), $t$ is the proper time of the oscillator which coincides with the
Schwarzschild coordinate $t$ because oscillators are located at fix $r=0$.
Since oscillators cannot emit Rindler photons, the Hamiltonian (\ref{b2a})
is not Hermitian.

We will consider evolution of the system as a function of the oscillator
proper time $t$. Schr\"{o}dinger equation for the system's state vector%
\begin{equation*}
i\hslash \frac{\partial }{\partial t}\left\vert \psi (t)\right\rangle =\hat{V%
}_{2}(t)\left\vert \psi (t)\right\rangle
\end{equation*}%
yields%
\begin{equation}
\left\vert \psi (t)\right\rangle =e^{\beta (t)\hat{\sigma}\hat{b}%
_{2}}\left\vert 0_{U}\right\rangle \left\vert A\right\rangle ,  \label{c1b}
\end{equation}%
where $\left\vert 0_{U}\right\rangle $ and $\left\vert A\right\rangle $ are
the initial state vectors of the field and the oscillator, and%
\begin{equation*}
\beta (t)=-\frac{ig}{\hslash }\int_{0}^{t}dt^{\prime }e^{-i\omega t^{\prime
}}\phi _{2}(t^{\prime })=-\frac{ig}{\hslash }\int_{0}^{t}dt^{\prime
}e^{i\left( \frac{c\Omega }{2r_{g}}-\omega \right) t^{\prime }}.
\end{equation*}%
Plug $\left\vert 0_{U}\right\rangle $ in Eq. (\ref{c1b}) gives%
\begin{equation}
\left\vert \psi (t)\right\rangle =\sqrt{1-\gamma ^{2}}e^{\beta (t)\hat{\sigma%
}\hat{b}_{2}}e^{\gamma \hat{b}_{1}^{\dagger }\hat{b}_{2}^{\dagger
}}\left\vert 0_{R}\right\rangle \left\vert A\right\rangle .
\end{equation}%
Using the Baker--Hausdorff formula $e^{\hat{A}}e^{\hat{B}}=e^{[\hat{A},\hat{B%
}]}e^{\hat{B}}e^{\hat{A}}$, we obtain%
\begin{equation*}
\left\vert \psi (t)\right\rangle =\sqrt{1-\gamma ^{2}}e^{\gamma \beta (t)%
\hat{\sigma}\hat{b}_{1}^{\dagger }}e^{\gamma \hat{b}_{1}^{\dagger }\hat{b}%
_{2}^{\dagger }}e^{\beta (t)\hat{\sigma}\hat{b}_{2}}\left\vert
0_{R}\right\rangle \left\vert A\right\rangle ,
\end{equation*}%
or%
\begin{equation}
\left\vert \psi (t)\right\rangle =e^{\gamma \beta (t)\hat{\sigma}\hat{b}%
_{1}^{\dagger }}\left\vert 0_{U}\right\rangle \left\vert A\right\rangle .
\label{c3a}
\end{equation}%
Equation (\ref{c3a}) shows that non-unitary field absorption at the
spacetime boundary yields generation of photons outside the BH event horizon
(into the Rindler mode $1$). Taking time derivative of Eq. (\ref{c3a}) leads
to the Schr\"{o}dinger equation with the interaction Hamiltonian 
\begin{equation*}
\hat{V}_{1}(t)=\gamma g\hat{\sigma}e^{-i\omega t}\phi _{1}^{\ast }(t)\hat{b}%
_{1}^{\dagger },
\end{equation*}%
where we used $\phi _{2}(t)=\phi _{1}^{\ast }(t)=\phi _{1}^{\ast
}(-T(t,0),X(t,0))$.

That is, the process looks like as if there is a mirror \textquotedblleft
image\textquotedblright\ of the oscillator located along the line $T=-\sqrt{%
1+X^{2}}$ (see Fig. \ref{Fig1}) which is coupled with the external mode $%
\phi _{1}$ with a reduced coupling constant $\gamma g$. The oscillator's
image produces field outside the event horizon which propagates away from
the BH. Such field is not thermal. E.g., if the oscillator is in a coherent
state, the generated field is coherent. The information stored in the
oscillators is recorded in the outgoing field.

BH radiation is not thermal because evolution of the field under the horizon
is described by the non-Hermitian Hamiltonian (\ref{b2a}). Indeed, if the
Hamiltonian would be Hermitian and depends only on $\hat{b}_{2}$ and $\hat{b}%
_{2}^{\dagger }$, the Heisenberg equation of motion for the operator $\hat{b}%
_{1}(t)$%
\begin{equation}
\frac{d\hat{b}_{1}(t)}{dt}=\frac{i}{\hslash }\left( \hat{H}^{\dagger }\hat{b}%
_{1}(t)-\hat{b}_{1}(t)\hat{H}\right) =\frac{i}{\hslash }\left( \hat{H}%
^{\dagger }(t)-\hat{H}(t)\right) \hat{b}_{1}  \label{c4b}
\end{equation}%
would yield $\hat{b}_{1}(t)=const$. That is field outside the BH event
horizon would not change. However, if $\hat{H}^{\dagger }\neq \hat{H}$, the
right-hand-side of Eq. (\ref{c4b}) is no longer zero and the external field
can be altered.

In the present model of BH evaporation the von Neumann entropy is preserved.
Namely, since evolution of the system is described by a Hamiltonian, the
system remains in a pure state and, thus, the net entropy remains equal to
zero. This is true even if the Hamiltonian is not Hermitian. For the latter,
the system's state vector should be normalized such that $\left\langle \psi
\right. \left\vert \psi \right\rangle =1$.

The toy model Hamiltonian (\ref{b2a}) explains why non-unitary absorption of
photons at the BH center alters radiation outside the BH. However, it does
not describe the system's dynamics correctly. The point is that,
non-Hermitian Hamiltonians don't preserve the expectation value of an
operator $\hat{Q}$ with which they commute. This is the reason why the norm
of the state vector is not conserved (in this case $\hat{Q}=1$). To
incorporate a conservation law $\left\langle \hat{Q}\right\rangle =const$
into the model, we must replace the non-Hermitian Hamiltonian $\hat{H}$ with
a constrained Hamiltonian \cite{Gross59,Amio75}%
\begin{equation}
\hat{H}-\lambda (t)\hat{Q},
\end{equation}%
where $\lambda (t)$ is a Lagrange multiplier whose value is to be chosen so
as to honor the constraint condition $\left\langle \hat{Q}\right\rangle
=const$.

We will impose a constraint that during BH evaporation the average energy is
conserved. Operators describing conserved quantities must commute with the
Hamiltonian. Such \textquotedblleft energy\textquotedblright\ operators
commuting with the Hamiltonian (\ref{b2a}) are%
\begin{equation*}
\hat{\sigma}^{\dagger }\hat{\sigma}-\hat{b}_{2}^{\dagger }\hat{b}_{2},\text{%
\ and }\hat{b}_{1}^{\dagger }\hat{b}_{1},
\end{equation*}%
and the constraints read 
\begin{equation}
\left\langle \hat{\sigma}^{\dagger }\hat{\sigma}-\hat{b}_{2}^{\dagger }\hat{b%
}_{2}\right\rangle =const\text{ \ and \ }\left\langle \hat{b}_{1}^{\dagger }%
\hat{b}_{1}\right\rangle =const.
\end{equation}%
The constrained interaction Hamiltonian is%
\begin{equation}
\hat{V}(t)=g\hat{\sigma}e^{-i\omega t}\phi _{2}(t)\hat{b}_{2}+i\hslash \hat{C%
}(t),  \label{IH}
\end{equation}%
where 
\begin{equation*}
\hat{C}(t)=\dot{\mu}_{1}(t)\hat{b}_{1}^{\dagger }\hat{b}_{1}+\dot{\mu}%
_{2}(t)\left( \hat{\sigma}^{\dagger }\hat{\sigma}-\hat{b}_{2}^{\dagger }\hat{%
b}_{2}\right) +\dot{\mu}_{3}(t)
\end{equation*}%
and the dot denotes derivative over $t$. The latter is introduced for
convenience. We assume that the resonance condition $\omega =c\Omega /2r_{g}$
is satisfied, which yields%
\begin{equation}
\hat{V}(t)=g\hat{\sigma}\hat{b}_{2}+i\hslash \hat{C}(t).  \label{CH}
\end{equation}%
The Lagrange multiplier $\dot{\mu}_{3}(t)$ takes into account the
normalization condition $\left\langle \psi \right. \left\vert \psi
\right\rangle =1$. For the present problem, Lagrange multipliers $\dot{\mu}%
_{1,2,3}(t)$ are real functions.

We assume that initially the oscillator is in a coherent state $\left\vert
A\right\rangle $ and the field is in the Unruh vacuum $\left\vert
0_{U}\right\rangle $. Schr\"{o}dinger equation with the constrained
Hamiltonian (\ref{CH}) yields (see Appendix A and B) 
\begin{equation}
\left\vert \psi (t)\right\rangle =N(t)e^{-\frac{i}{\hslash }\gamma gAe^{\mu
_{1}(t)}t\hat{b}_{1}^{\dagger }}e^{e^{\mu _{1}(t)-\mu _{2}(t)}\gamma \hat{b}%
_{1}^{\dagger }\hat{b}_{2}^{\dagger }}\left\vert 0_{R}\right\rangle
\left\vert e^{\mu _{2}(t)}A\right\rangle ,  \label{C0}
\end{equation}%
where $N(t)$ is a normalization factor and the Lagrange multipliers are
obtained from the constraint equations%
\begin{equation}
e^{2\mu _{2}}A^{2}-\frac{\tilde{\gamma}^{2}}{1-\tilde{\gamma}^{2}}-\frac{%
e^{2\mu _{1}}\tilde{\gamma}^{2}\left( \gamma \Lambda t\right) ^{2}}{\left( 1-%
\tilde{\gamma}^{2}\right) ^{2}}=A^{2}-\frac{\gamma ^{2}}{1-\gamma ^{2}},
\label{C1}
\end{equation}%
\begin{equation}
\frac{\tilde{\gamma}^{2}}{1-\tilde{\gamma}^{2}}+\frac{e^{2\mu _{1}}\left(
\gamma \Lambda t\right) ^{2}}{\left( 1-\tilde{\gamma}^{2}\right) ^{2}}=\frac{%
\gamma ^{2}}{1-\gamma ^{2}},  \label{C2}
\end{equation}%
where $\tilde{\gamma}=e^{\mu _{1}-\mu _{2}}\gamma $ and $\Lambda =gA/\hslash 
$ is the Rabi frequency. For $t\rightarrow \infty $, Eqs. (\ref{C0})-(\ref%
{C2}) give%
\begin{equation}
\left\vert \psi (\infty )\right\rangle =Ne^{-\frac{i\gamma }{\sqrt{1-\gamma
^{2}}}\hat{b}_{1}^{\dagger }}\left\vert 0_{R}\right\rangle \left\vert
A_{\infty }\right\rangle ,  \label{C4}
\end{equation}%
where 
\begin{equation}
A_{\infty }^{2}=A^{2}-\frac{\gamma ^{2}}{1-\gamma ^{2}}
\end{equation}%
is the mean number of oscillator excitations in the final state. The present
model of the spacetime boundary is self-consistent if the oscillators absorb
all ingoing photons, which implies $A_{\infty }^{2}>0$. Otherwise, photon
flux through the boundary would be nonzero.

Eq. (\ref{C4}) shows that the final state of the field is the Rindler vacuum
for photons $\hat{b}_{2}$ and a coherent state for photons $\hat{b}_{1}$
with the average photon number $\gamma ^{2}/(1-\gamma ^{2})$. The oscillator
remains in the coherent state, but the oscillator's mean excitation number
is reduced by an amount $\gamma ^{2}/(1-\gamma ^{2})$ due to absorption of
all $\hat{b}_{2}$ photons.

For in our model $\left\langle \hat{b}_{1}^{\dagger }\hat{b}%
_{1}\right\rangle =const$, the radiation power of an evaporating BH is given
by the Hawking's formula, but photon statistics is not thermal and the
outgoing radiation carries information about the BH interior. In particular,
coherent oscillations of the BH interior lead to a coherent outgoing
radiation.

In the limit $\gamma \ll 1$, Eqs. (\ref{C1}) and (\ref{C2}) can be solved
analytically yielding the following expression for the system's state vector
as a function of $t$%
\begin{equation}
\left\vert \psi (t)\right\rangle =N(t)e^{-\frac{i\gamma \Lambda t\hat{b}%
_{1}^{\dagger }}{\sqrt{1+\Lambda ^{2}t^{2}}}}e^{\frac{\gamma \hat{b}%
_{1}^{\dagger }\hat{b}_{2}^{\dagger }}{\sqrt{1+\Lambda ^{2}t^{2}}}%
}\left\vert 0_{R}\right\rangle \left\vert A(t)\right\rangle ,  \label{C5}
\end{equation}%
where 
\begin{equation*}
A^{2}(t)=A^{2}-\frac{\Lambda ^{2}t^{2}}{1+\Lambda ^{2}t^{2}}\gamma ^{2}.
\end{equation*}%
According to Eq. (\ref{C5}), initial thermal Hawking radiation evolves into
the coherent state $e^{-i\gamma \hat{b}_{1}^{\dagger }}\left\vert
0_{R}\right\rangle $ on a time scale $1/\Lambda $, while the oscillator's
energy (BH mass) decreases as $\hslash \omega A^{2}(t)$.

\section{Insights from quantum gravity models}

Here we show that present mechanism of nonthermal emission of evaporating
BHs holds for an effective metric obtained in quantum gravity models. Most
of such models suggest that the classical singularity at $r=0$ should be
replaced by a regular timelike boundary. To be specific, we consider an
effective BH metric obtained from scale-dependent effective average action
which takes into account the effect of all loops \cite{Wett93,Reut94,Reut98}%
. As a function of this scale, the effective average action satisfies a
renormalization group equation yielding the effective metric \cite{Bona00}%
\begin{equation}
ds^{2}=f(r)c^{2}dt^{2}-\frac{1}{f(r)}dr^{2}-r^{2}\left( d\theta ^{2}+\sin
^{2}\theta d\varphi ^{2}\right) ,  \label{q4a}
\end{equation}%
where 
\begin{equation}
f(r)=1-\frac{r_{g}}{r}\frac{1}{1+\frac{\bar{\omega}r_{g}^{2}}{r^{2}}},
\label{FR}
\end{equation}%
and $\bar{\omega}>0$ is a constant that involves the quantum gravity
correction to the BH geometry coming from the renormalization group approach.

The metric (\ref{q4a}) is regular at $r=0$ and has two horizons which can be
found by setting $f(r)=0$ in Eq. (\ref{FR}). The position of the outer and
inner horizons is%
\begin{equation*}
r_{\pm }=\frac{r_{g}}{2}\left( 1\pm \sqrt{1-4\bar{\omega}}\right) .
\end{equation*}%
In terms of $r_{\pm }$, one can write%
\begin{equation*}
\frac{1}{f(r)}=1+\frac{r_{g}r}{(r-r_{-})(r-r_{+})}.
\end{equation*}

A massless scalar field $\phi $ obeys the covariant wave equation 
\begin{equation}
\frac{1}{\sqrt{-g}}\frac{\partial }{\partial x^{\mu }}\left( \sqrt{-g}g^{\mu
\nu }\frac{\partial \phi }{\partial x^{\nu }}\right) =0,  \label{w3}
\end{equation}%
where $g^{\mu \nu }$ is the spacetime metric given by the interval (\ref{q4a}%
), namely%
\begin{equation*}
g^{tt}=\frac{1}{f(r)},\quad g^{rr}=-f(r).
\end{equation*}%
For the truncated 1+1 dimensional spacetime $\sqrt{-g}=1$, and the wave
equation (\ref{w3}) reduces to 
\begin{equation}
\frac{1}{c^{2}}\frac{\partial ^{2}\phi }{\partial t^{2}}-f(r)\frac{\partial 
}{\partial r}\left( f(r)\frac{\partial \phi }{\partial r}\right) =0.
\label{w3a}
\end{equation}%
Solutions of Eq. (\ref{w3a}) read%
\begin{equation}
\phi _{\nu }(t,r)=e^{-i\nu \left[ t\pm \frac{r}{c}\mp \chi (r)\right] },
\label{w4}
\end{equation}%
where%
\begin{equation*}
\chi (r)=\frac{r_{-}\ln |r-r_{-}|-r_{+}\ln |r-r_{+}|}{c\sqrt{1-4\bar{\omega}}%
}.
\end{equation*}

Using Eq. (\ref{w4}), one can construct mode functions analogous to the
Rindler modes (\ref{R1a}) and (\ref{R1b}) in the Schwarzschild coordinates,
namely, 
\begin{equation}
\phi _{1\nu }(t,r)=e^{-i\nu \left[ t-\frac{r}{c}+\chi (r)\right] }\theta
(r-r_{+}),  \label{w5a}
\end{equation}%
\begin{equation}
\phi _{2\nu }(t,r)=e^{i\nu \left[ t-\frac{r}{c}+\chi (r)\right] }\theta
(r_{+}-r)\theta (r-r_{-}),  \label{w5b}
\end{equation}%
where $\nu >0$. For $r_{-}=0$, the mode functions (\ref{w5a}) and (\ref{w5b}%
) reduce to Eqs. (\ref{R1a}) and (\ref{R1b}) with $\Omega =2r_{g}\nu /c$.

Eqs. (\ref{w5a}) and (\ref{w5b}) show that if an observer is held fixed
outside the outer event horizon at a constant $r>r_{+}$, then at the
observer's location the non-zero Rindler modes $\phi _{1\nu }$ oscillate as $%
\phi _{1\nu }\propto e^{-i\nu t}$, where $t$ is the observer's proper time.
That is, from the observer's perspective, the Rindler photons $\phi _{1\nu }$
behave as if they have positive frequency $\nu $. However, if a hypothetical
observer is located at fixed $r_{-}<r<r_{+}$, the non-zero Rindler mode $%
\phi _{2\nu }$ oscillates as a function of the proper time $t$ as $\phi
_{2\nu }\propto e^{i\nu t}$. That is, from the observer's perspective, the
Rindler photons $\phi _{2\nu }$ behave as if they have negative frequency $%
-\nu $. Absorption of photons $\phi _{2\nu }$ decreases energy (mass) of the
BH, leading to BH evaporation.

Photons falling into the BH from BH exterior are described by the mode
functions%
\begin{equation}
\phi _{3\nu }(t,r)=e^{-i\nu \left[ t+\frac{r}{c}-\chi (r)\right]
}-e^{i\varphi _{0}}e^{-i\nu \left[ t-\frac{r}{c}+\chi (r)\right] }\theta
(r_{-}-r),  \label{w5c}
\end{equation}%
where the last term describes a wave reflected from the timelike spacetime
boundary $r=0$, and $\varphi _{0}$ is a phase shift introduced to satisfy
the reflective boundary condition, e.g., $\left. \partial \phi _{3\nu
}/\partial r\right\vert _{r=0}=0$. From the perspective of an observer held
fixed at $r=const$ the mode functions $\phi _{3\nu }$ have positive
frequency. Thus, absorption of such photons increases the BH mass.

\begin{figure}[h]
\begin{center}
\epsfig{figure=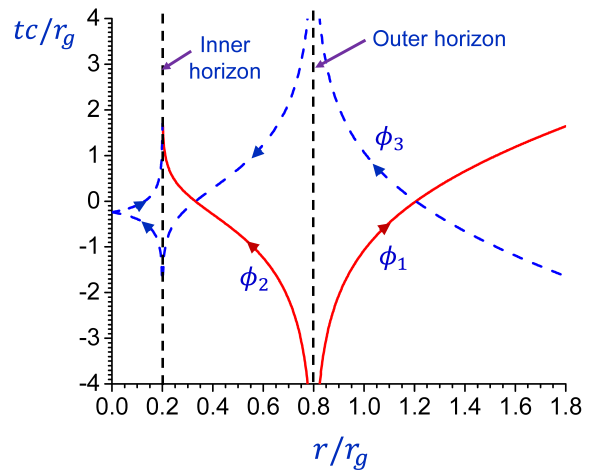, angle=0, width=9cm}
\end{center}
\caption{Light rays of photons $\protect\phi _{1\protect\nu }$, $\protect%
\phi _{2\protect\nu }$ (solid line) and $\protect\phi _{3\protect\nu }$
(dash line) in the metric (\protect\ref{q4a}) for $r_{-}=0.2r_{g}$ and $%
r_{+}=0.8r_{g}$.}
\label{Fig4}
\end{figure}

In Fig. \ref{Fig4} we plot light rays of photons (\ref{w5a}), (\ref{w5b})
and (\ref{w5c}) in the Schwarzschild coordinates. The figure shows that the
negative ($\phi _{2\nu }$) and positive ($\phi _{1\nu }$) frequency Rindler
photons are generated at the outer horizon. These photons are produced in
pairs and are entangled. Photons $\phi _{1\nu }$ carry energy away from BH,
while the negative energy photons $\phi _{2\nu }$ propagate toward the BH
center and are absorbed at the inner horizon. The positive energy photons $%
\phi _{3\nu }$ carry energy into the BH from the BH exterior. They cross
both outer and inner horizons, and after reflection from the BH center are
absorbed at the inner horizon. In the region $r_{-}<r<r_{+}$, the coordinate 
$r$ plays the role of time for particles which move unidirectionally along
the $r$ coordinate in this region. For $r<r_{-}$ and $r>r_{+}$ the particles
move unidirectionally along the $t$ coordinate.

Spacetime described by the metric (\ref{q4a}) is non-singular and matter
does not disappear. Figure \ref{Fig4} shows that matter and energy
(infalling photons $\phi _{3\nu }$) are concentrated in the vicinity of the
inner horizon. Since Rindler photons $\phi _{2\nu }$ can only be annihilated
and not created at the inner horizon, the non-unitary absorption of the
Rindler photons $\phi _{2\nu }$ at the inner horizon, combined with the
entanglement of photon pairs $\phi _{1\nu }$ and $\phi _{2\nu }$ generated
at the outer horizon, leads to nonthermal outgoing radiation that carries
information about the BH interior. One can model this process by the same
Hamiltonian (\ref{IH}) of the previous section, but now the oscillators
absorbing the ingoing photons $\phi _{2\nu }$ follow the worldline of the
inner horizon and can be a model of matter rather than gravitational field.

The picture becomes more intuitive if we describe BH evaporation in terms of
particles and antiparticles that can annihilate with each other. In this
picture, particle ($\phi _{1\nu }$) and antiparticle ($\phi _{2\nu }$) are
generated as entangled pairs at the outer horizon. The particles $\phi
_{1\nu }$ carry energy away from BH. The antiparticles move towards BH
center and at the inner horizon annihilate with particles $\phi _{3\nu }$
which have been accumulated at the inner horizon during BH formation. Due to
entanglement between $\phi _{1\nu }$ and $\phi _{2\nu }$, the information
about state of particles $\phi _{3\nu }$ is recorded into the outgoing flux
of particles $\phi _{1\nu }$.

\section{Summary and Discussion}

Evaporation of a classical Schwarzschild BH is caused by creation of
entangled particle-antiparticle pairs (Rindler photons in the present
discussion) near the event horizon, with one carrying positive energy to
infinity and the other carrying negative energy into the BH. This is the
mechanism of Hawking radiation. Absorption of the negative energy photons at
the center of the classical BH reduces the BH mass.

Here we argue that previous models of Hawking radiation are lacking an
important ingredient. Namely, the process of photon absorption at the BH
center must be properly described quantum mechanically by constructing a
Hamiltonian. Since under the BH event horizon, light can propagate only
towards the BH center, the symmetry between absorption and emission is
broken. Namely, BH center can only absorb photons, but do not emit. As a
result, the Hamiltonian describing BH evaporation is not Hermitian.

To describe absorption of photons at the BH center, we assume that the
latter consists of harmonic oscillators which absorb the ingoing radiation,
but do not emit. In our model, the oscillators follow the worldline of the
BH center, rather than geodesics, and carry information about the BH
interior.

We show that due to entanglement between photons moving inside and outside
the BH event horizon, the non-unitary absorption of the radiation under the
horizon alters the state of the field outside the BH. As a consequence,
radiation produced by the evaporating BH is not thermal and carries
information about the BH interior. After the BH has evaporated, the
information is recorded in the remaining non-thermal field. Since evolution
is governed by a Hamiltonian, the state of the system remains pure and
during BH evaporation the von Neumann entropy is preserved. In our model we
impose a constraint that energy is conserved during BH evaporation. As a
consequence, our model yields that luminosity of an evaporating BH coincides
with that for Hawking radiation.

Erasing information at the BH center produced by photon absorption is a
non-unitary process which leads to a change of the field outside the
horizon. This is somewhat analogous to the quantum eraser experiments in
which the interference pattern can be destroyed or restored by manipulating
entangled photon partners \cite{Scul82,Kim00,Walb02}. In these experiments,
after two entangled photons are created, each is directed into different
section of the apparatus and an interference pattern for one of them is
examined. A measurement done on the entangled partner to learn about the
photon path influences the interference pattern.

Similarly to BH evaporation, non-unitarity of the measurement process alters
the state of the entangled partner. However, the state vector collapse
brought about by a measurement is a probabilistic and discontinuous change,
while BH evaporation is a deterministic, continuous time evolution of an
isolated system that obeys the Schr\"{o}dinger equation.

Our findings show that quantum mechanical evolution, governed by the Schr%
\"{o}dinger equation, allows information to leak out from the BH. This is
the case because BH center breaks the emission-absorption symmetry and
photons external to the horizon are entangled with those inside it. Such
entanglement is an inherent property of the field for evaporating BHs.

We also show that present mechanism of nonthermal emission of evaporating
BHs holds for spacetimes obtained in quantum gravity models in which the
classical singularity at $r=0$ is replaced by a regular timelike boundary.
For such spacetimes the metric has an inner and outer horizons, and matter
does not disappear. Instead, particles are accumulated in the vicinity of
the inner horizon. For this spacetime, the entangled particle-antiparticle
pairs are generated at the outer horizon. The generated particles carry
energy away from BH, while antiparticles move towards the BH center and
annihilate at the inner horizon with particles that form the BH interior.
Due to entanglement of the particle-antiparticle pairs produced at the outer
horizon, the information about the BH interior is recorded in the outgoing
particle flux.

One should mention that if our findings are correct, and radiation of
evaporating BHs is nonthermal, the Bekenstein-Hawking formula \cite%
{Beke72,Hawk76b} does not describe the BH entropy. Recall that the latter
formula assumes thermal BH emission with the Hawking temperature.

Our results demonstrate that quantum mechanics works in an exotic spacetime
geometry of a BH. However, BHs might have only a mathematical significance.
The point is that there is an evidence that general relativity is ruled out
by gravitational waves detection experiments in favor of the vector theory
of gravity \cite{Svid21b}. The latter theory \cite{Svid17,Svid19a} agrees
with all available tests of gravity, including detection of gravitational
waves and observations of supermassive objects at galactic centers \cite%
{Svid07,Svid21b}. In addition, vector gravity predicts no BHs and yields the
measured value of the cosmological constant \cite{Plan14} with no free
parameters \cite{Svid17,Svid19a}.

\begin{acknowledgements}
This work was supported by the Air Force Office of Scientific Research
(Grant No. FA9550-20-1-0366 DEF), the Robert A. Welch Foundation (Grant No. A-1261), and  the
National Science Foundation (Grant No. PHY-2013771).
\end{acknowledgements}

\appendix

\section{Operator identities and expectation values}

Operators of Rindler photons $\hat{b}_{1}$ and $\hat{b}_{2}$ obey bosonic
commutation relations%
\begin{equation*}
\lbrack \hat{b}_{1},\hat{b}_{1}^{\dagger }]=1,\quad \lbrack \hat{b}_{2},\hat{%
b}_{2}^{\dagger }]=1,
\end{equation*}%
and all other commutators are equal to zero. First we prove an operator
identity%
\begin{equation}
\hat{b}_{2}e^{\gamma \hat{b}_{1}^{\dagger }\hat{b}_{2}^{\dagger }}=e^{\gamma 
\hat{b}_{1}^{\dagger }\hat{b}_{2}^{\dagger }}\hat{b}_{2}+\gamma \hat{b}%
_{1}^{\dagger }e^{\gamma \hat{b}_{1}^{\dagger }\hat{b}_{2}^{\dagger }},
\label{v0}
\end{equation}%
where $\gamma $ is a complex number. Introducing operator 
\begin{equation*}
\hat{F}(\gamma )=\hat{b}_{2}e^{\gamma \hat{b}_{1}^{\dagger }\hat{b}%
_{2}^{\dagger }}-e^{\gamma \hat{b}_{1}^{\dagger }\hat{b}_{2}^{\dagger }}\hat{%
b}_{2},
\end{equation*}%
we have%
\begin{equation*}
\frac{d\hat{F}(\gamma )}{d\gamma }=\hat{b}_{2}\hat{b}_{1}^{\dagger }\hat{b}%
_{2}^{\dagger }e^{\gamma \hat{b}_{1}^{\dagger }\hat{b}_{2}^{\dagger }}-\hat{b%
}_{1}^{\dagger }\hat{b}_{2}^{\dagger }e^{\gamma \hat{b}_{1}^{\dagger }\hat{b}%
_{2}^{\dagger }}\hat{b}_{2}=\hat{b}_{1}^{\dagger }\hat{b}_{2}^{\dagger }\hat{%
F}(\gamma )+\hat{b}_{1}^{\dagger }e^{\gamma \hat{b}_{1}^{\dagger }\hat{b}%
_{2}^{\dagger }}.
\end{equation*}%
Solution of this differential equation, subject to the condition $\hat{F}%
(0)=0,$ is 
\begin{equation*}
\hat{F}(\gamma )=\gamma \hat{b}_{1}^{\dagger }e^{\gamma \hat{b}_{1}^{\dagger
}\hat{b}_{2}^{\dagger }},
\end{equation*}%
which proves the identity (\ref{v0}).

Next we prove an identity%
\begin{equation}
e^{\lambda \hat{b}_{2}^{\dagger }\hat{b}_{2}}\hat{b}_{2}^{\dagger
}=e^{\lambda }\hat{b}_{2}^{\dagger }e^{\lambda \hat{b}_{2}^{\dagger }\hat{b}%
_{2}},  \label{v0a}
\end{equation}%
where $\lambda $ is a complex number. Introducing operator 
\begin{equation*}
\hat{F}(\lambda )=e^{\lambda \hat{b}_{2}^{\dagger }\hat{b}_{2}}\hat{b}%
_{2}^{\dagger }-\hat{b}_{2}^{\dagger }e^{\lambda \hat{b}_{2}^{\dagger }\hat{b%
}_{2}},
\end{equation*}%
we have%
\begin{equation*}
\frac{d\hat{F}(\lambda )}{d\lambda }=\hat{b}_{2}^{\dagger }\hat{b}%
_{2}e^{\lambda \hat{b}_{2}^{\dagger }\hat{b}_{2}}\hat{b}_{2}^{\dagger }-\hat{%
b}_{2}^{\dagger }\hat{b}_{2}^{\dagger }\hat{b}_{2}e^{\lambda \hat{b}%
_{2}^{\dagger }\hat{b}_{2}}=\hat{b}_{2}^{\dagger }\hat{b}_{2}\hat{F}(\lambda
)+\hat{b}_{2}^{\dagger }e^{\lambda \hat{b}_{2}^{\dagger }\hat{b}_{2}}.
\end{equation*}%
Solution of this differential equation, subject to the condition $\hat{F}%
(0)=0,$ is 
\begin{equation*}
\hat{F}(\lambda )=\left( e^{\lambda }-1\right) \hat{b}_{2}^{\dagger
}e^{\lambda \hat{b}_{2}^{\dagger }\hat{b}_{2}},
\end{equation*}%
which proves the identity (\ref{v0a}).

Next we prove an identity%
\begin{equation}
e^{\lambda \hat{b}_{2}^{\dagger }\hat{b}_{2}}e^{\gamma \hat{b}_{1}^{\dagger }%
\hat{b}_{2}^{\dagger }}=e^{e^{\lambda }\gamma \hat{b}_{1}^{\dagger }\hat{b}%
_{2}^{\dagger }}e^{\lambda \hat{b}_{2}^{\dagger }\hat{b}_{2}}.  \label{v3}
\end{equation}%
Introducing operator 
\begin{equation*}
\hat{F}(\lambda )=e^{\lambda \hat{b}_{2}^{\dagger }\hat{b}_{2}}e^{\gamma 
\hat{b}_{1}^{\dagger }\hat{b}_{2}^{\dagger }}e^{-\lambda \hat{b}%
_{2}^{\dagger }\hat{b}_{2}},
\end{equation*}%
and taking derivative over $\lambda $, we have 
\begin{equation*}
\frac{d\hat{F}(\lambda )}{d\lambda }=e^{\lambda \hat{b}_{2}^{\dagger }\hat{b}%
_{2}}\hat{b}_{2}^{\dagger }\hat{b}_{2}e^{\gamma \hat{b}_{1}^{\dagger }\hat{b}%
_{2}^{\dagger }}e^{-\lambda \hat{b}_{2}^{\dagger }\hat{b}_{2}}-e^{\lambda 
\hat{b}_{2}^{\dagger }\hat{b}_{2}}e^{\gamma \hat{b}_{1}^{\dagger }\hat{b}%
_{2}^{\dagger }}\hat{b}_{2}^{\dagger }\hat{b}_{2}e^{-\lambda \hat{b}%
_{2}^{\dagger }\hat{b}_{2}}.
\end{equation*}%
Taking into account identities (\ref{v0}) and (\ref{v0a}), we obtain 
\begin{equation*}
\frac{d\hat{F}(\lambda )}{d\lambda }=\gamma \hat{b}_{1}^{\dagger }e^{\lambda 
\hat{b}_{2}^{\dagger }\hat{b}_{2}}\hat{b}_{2}^{\dagger }e^{\gamma \hat{b}%
_{1}^{\dagger }\hat{b}_{2}^{\dagger }}e^{-\lambda \hat{b}_{2}^{\dagger }\hat{%
b}_{2}}=\gamma e^{\lambda }\hat{b}_{1}^{\dagger }\hat{b}_{2}^{\dagger }\hat{F%
}(\lambda ).
\end{equation*}%
Solution of this differential equation, subject to the condition $\hat{F}%
(0)=e^{\gamma \hat{b}_{1}^{\dagger }\hat{b}_{2}^{\dagger }},$ is 
\begin{equation*}
\hat{F}(\lambda )=e^{e^{\lambda }\gamma \hat{b}_{1}^{\dagger }\hat{b}%
_{2}^{\dagger }},
\end{equation*}%
which proves the identity (\ref{v3}).

Next we calculate a matrix element $\left\langle \psi |\psi \right\rangle $,
where state vector $\left\vert \psi \right\rangle $ is 
\begin{equation}
\left\vert \psi \right\rangle =\sqrt{1-\gamma ^{2}}e^{\beta \hat{b}%
_{1}^{\dagger }}e^{\gamma \hat{b}_{1}^{\dagger }\hat{b}_{2}^{\dagger
}}\left\vert 0_{R}\right\rangle ,  \label{v1}
\end{equation}%
$\left\vert 0_{R}\right\rangle $ stands for the Rindler vacuum, $\beta $ is
a complex number and $\gamma $ is a real number. The state vector (\ref{v1})
can be written as%
\begin{equation*}
\left\vert \psi \right\rangle =e^{\beta \hat{b}_{1}^{\dagger }}\left\vert
0_{M}\right\rangle ,
\end{equation*}%
where%
\begin{equation}
\left\vert 0_{M}\right\rangle =\sqrt{1-\gamma ^{2}}e^{\gamma \hat{b}%
_{1}^{\dagger }\hat{b}_{2}^{\dagger }}\left\vert 0_{R}\right\rangle
\label{MV}
\end{equation}%
is the Minkowski vacuum. Using a relation between operators of the Rindler
photons $\hat{b}_{1,2}$ and the Unruh-Minkowski photons $\hat{a}_{1,2}$ \cite%
{Svid21a}%
\begin{equation*}
\hat{b}_{1}^{\dagger }=\frac{\hat{a}_{1}^{\dagger }+\gamma \hat{a}_{2}}{%
\sqrt{1-\gamma ^{2}}},
\end{equation*}%
and the property $\hat{a}_{1,2}\left\vert 0_{M}\right\rangle =0$, we obtain%
\begin{equation*}
\left\vert \psi \right\rangle =e^{\frac{\beta \hat{a}_{1}^{\dagger }}{\sqrt{%
1-\gamma ^{2}}}}\left\vert 0_{M}\right\rangle .
\end{equation*}%
Taking into account that%
\begin{equation*}
e^{\alpha \hat{a}_{1}^{\dagger }}\left\vert 0_{M}\right\rangle =e^{\frac{%
|\alpha |^{2}}{2}}\left\vert \alpha 0\right\rangle ,
\end{equation*}%
where $\left\vert \alpha 0\right\rangle $ stands for a coherent state $%
\left\vert \alpha \right\rangle $ for the Unruh-Minkowski photons $\hat{a}%
_{1}$ and the vacuum state for the Unruh-Minkowski photons $\hat{a}_{2}$, we
find%
\begin{equation}
\left\vert \psi \right\rangle =e^{\frac{|\beta |^{2}}{2\left( 1-\gamma
^{2}\right) }}\left\vert \alpha 0\right\rangle ,  \label{s1}
\end{equation}%
where $\alpha =\beta /\sqrt{1-\gamma ^{2}}$. Therefore%
\begin{equation}
\left\langle \psi |\psi \right\rangle =e^{\frac{|\beta |^{2}}{1-\gamma ^{2}}%
}.  \label{v2}
\end{equation}

Next we calculate the average number of Rindler photons $\hat{b}_{1}$ in the
state $\left\vert \psi \right\rangle $, that is $\left\langle \hat{b}%
_{1}^{\dagger }\hat{b}_{1}\right\rangle \equiv \left\langle \psi \right\vert 
\hat{b}_{1}^{\dagger }\hat{b}_{1}\left\vert \psi \right\rangle /\left\langle
\psi |\psi \right\rangle $. Taking derivative of Eq. (\ref{v2}) with respect
to $\beta $ and $\beta ^{\ast }$, and using Eq. (\ref{v1}), we have%
\begin{equation*}
\left\langle \psi \right\vert \hat{b}_{1}\hat{b}_{1}^{\dagger }\left\vert
\psi \right\rangle =\frac{\partial ^{2}}{\partial \beta \partial \beta
^{\ast }}e^{\frac{|\beta |^{2}}{1-\gamma ^{2}}}=\frac{1-\gamma
^{2}+\left\vert \beta \right\vert ^{2}}{\left( 1-\gamma ^{2}\right) ^{2}}e^{%
\frac{|\beta |^{2}}{1-\gamma ^{2}}}.
\end{equation*}%
Therefore%
\begin{equation}
\left\langle \hat{b}_{1}^{\dagger }\hat{b}_{1}\right\rangle =\frac{%
\left\langle \psi \right\vert \hat{b}_{1}\hat{b}_{1}^{\dagger }\left\vert
\psi \right\rangle }{\left\langle \psi |\psi \right\rangle }-1=\frac{\gamma
^{2}}{1-\gamma ^{2}}+\frac{\left\vert \beta \right\vert ^{2}}{\left(
1-\gamma ^{2}\right) ^{2}}.  \label{s3}
\end{equation}

To find $\left\langle \hat{b}_{2}^{\dagger }\hat{b}_{2}\right\rangle $ we
use the relations between operators of the Rindler photons $\hat{b}_{1,2}$
and the Unruh-Minkowski photons $\hat{a}_{1,2}$ \cite{Svid21a}%
\begin{equation*}
\hat{b}_{2}=\frac{\hat{a}_{2}+\gamma \hat{a}_{1}^{\dagger }}{\sqrt{1-\gamma
^{2}}},\quad \hat{b}_{2}^{\dagger }=\frac{\hat{a}_{2}^{\dagger }+\gamma \hat{%
a}_{1}}{\sqrt{1-\gamma ^{2}}},
\end{equation*}%
which yield%
\begin{equation*}
\hat{b}_{2}^{\dagger }\hat{b}_{2}=\frac{1}{1-\gamma ^{2}}\left( \hat{a}%
_{2}^{\dagger }\hat{a}_{2}+\gamma ^{2}\hat{a}_{1}\hat{a}_{1}^{\dagger
}+\gamma \hat{a}_{1}\hat{a}_{2}+\gamma \hat{a}_{2}^{\dagger }\hat{a}%
_{1}^{\dagger }\right) .
\end{equation*}%
Using Eq. (\ref{s1}), we obtain%
\begin{equation*}
\left\langle \psi \right\vert \hat{b}_{2}^{\dagger }\hat{b}_{2}\left\vert
\psi \right\rangle =\frac{\gamma ^{2}\left( 1+|\alpha |^{2}\right) }{%
1-\gamma ^{2}}e^{\frac{|\beta |^{2}}{1-\gamma ^{2}}},
\end{equation*}%
where we took into account that $\left\langle \alpha 0\right\vert \hat{a}_{1}%
\hat{a}_{1}^{\dagger }\left\vert \alpha 0\right\rangle =1+|\alpha |^{2}$. As
a result, 
\begin{equation}
\left\langle \hat{b}_{2}^{\dagger }\hat{b}_{2}\right\rangle =\frac{%
\left\langle \psi \right\vert \hat{b}_{2}^{\dagger }\hat{b}_{2}\left\vert
\psi \right\rangle }{\left\langle \psi |\psi \right\rangle }=\frac{\gamma
^{2}}{1-\gamma ^{2}}+\frac{\gamma ^{2}|\beta |^{2}}{\left( 1-\gamma
^{2}\right) ^{2}}.  \label{s4}
\end{equation}

\section{State vector evolution during black hole evaporation}

For our model of black hole evaporation, the constrained interaction
Hamiltonian is%
\begin{equation*}
\hat{V}(t)=g\hat{\sigma}\hat{b}_{2}+i\hslash \dot{\mu}_{1}(t)\hat{b}%
_{1}^{\dagger }\hat{b}_{1}+i\hslash \dot{\mu}_{2}(t)\left( \hat{\sigma}%
^{\dagger }\hat{\sigma}-\hat{b}_{2}^{\dagger }\hat{b}_{2}\right) +i\hslash 
\dot{\mu}_{3}(t),
\end{equation*}%
where functions $\mu _{1,2,3}(t)$ are real, and the oscillator's lowering
and raising operators $\hat{\sigma}$ and $\hat{\sigma}^{\dagger }$ obey the
same bosonic commutation relations as the operators of Rindler photons.
Schrodinger equation for the evolution of the field state vector%
\begin{equation*}
i\hslash \frac{\partial }{\partial t}\left\vert \psi (t)\right\rangle =\hat{V%
}(t)\left\vert \psi (t)\right\rangle
\end{equation*}%
yields%
\begin{equation}
\left\vert \psi (t)\right\rangle =e^{\mu _{3}}e^{\mu _{1}\hat{b}%
_{1}^{\dagger }\hat{b}_{1}+\mu _{2}\left( \hat{\sigma}^{\dagger }\hat{\sigma}%
-\hat{b}_{2}^{\dagger }\hat{b}_{2}\right) -\frac{i}{\hslash }gt\hat{\sigma}%
\hat{b}_{2}}\left\vert 0_{M}\right\rangle \left\vert A\right\rangle ,
\label{s2}
\end{equation}%
where $\left\vert 0_{M}\right\rangle $ and $\left\vert A\right\rangle $ are
the initial state vectors of the field and the oscillator respectively. We
assume that the latter is a coherent state $\left\vert A\right\rangle $,
where $A$ is real, and the former is the Minkowski vacuum $\left\vert
0_{M}\right\rangle $. Recall that Unruh vacuum coincides with the Minkowski
vacuum for the right-moving photons.

Taking into account that $\hat{\sigma}\hat{b}_{2}$ commutes with $\hat{b}%
_{1}^{\dagger }\hat{b}_{1}$ and $\hat{\sigma}^{\dagger }\hat{\sigma}-\hat{b}%
_{2}^{\dagger }\hat{b}_{2}$, and plugging $\left\vert 0_{M}\right\rangle $
from Eq. (\ref{MV}) in Eq. (\ref{s2}), we obtain%
\begin{equation*}
\left\vert \psi (t)\right\rangle =
\end{equation*}%
\begin{equation*}
\sqrt{1-\gamma ^{2}}e^{\mu _{3}}e^{-\frac{i}{\hslash }gt\hat{\sigma}\hat{b}%
_{2}}e^{\mu _{1}\hat{b}_{1}^{\dagger }\hat{b}_{1}+\mu _{2}\left( \hat{\sigma}%
^{\dagger }\hat{\sigma}-\hat{b}_{2}^{\dagger }\hat{b}_{2}\right) }e^{\gamma 
\hat{b}_{1}^{\dagger }\hat{b}_{2}^{\dagger }}\left\vert 0_{R}\right\rangle
\left\vert A\right\rangle .
\end{equation*}%
Using identity (\ref{v3}), we have%
\begin{equation*}
\left\vert \psi (t)\right\rangle =\sqrt{1-\gamma ^{2}}e^{\mu _{3}}e^{-\frac{i%
}{\hslash }gt\hat{\sigma}\hat{b}_{2}}e^{e^{\mu _{1}-\mu _{2}}\gamma \hat{b}%
_{1}^{\dagger }\hat{b}_{2}^{\dagger }}e^{\mu _{2}\hat{\sigma}^{\dagger }\hat{%
\sigma}}\left\vert 0_{R}\right\rangle \left\vert A\right\rangle .
\end{equation*}

Taking into account that 
\begin{equation*}
e^{\mu _{2}\hat{\sigma}^{\dagger }\hat{\sigma}}\left\vert A\right\rangle =e^{%
\frac{|A|^{2}}{2}\left( e^{2\mu _{2}}-1\right) }\left\vert e^{\mu
_{2}}A\right\rangle ,
\end{equation*}%
we find%
\begin{equation*}
\left\vert \psi (t)\right\rangle =
\end{equation*}%
\begin{equation*}
\sqrt{1-\gamma ^{2}}e^{\mu _{3}+\frac{|A|^{2}}{2}\left( e^{2\mu
_{2}}-1\right) }e^{-\frac{i}{\hslash }gt\hat{\sigma}\hat{b}_{2}}e^{e^{\mu
_{1}-\mu _{2}}\gamma \hat{b}_{1}^{\dagger }\hat{b}_{2}^{\dagger }}\left\vert
0_{R}\right\rangle \left\vert e^{\mu _{2}}A\right\rangle .
\end{equation*}%
Since the initial state of the oscillator is the coherent state $\left\vert
A\right\rangle $, and $\hat{\sigma}\left\vert A\right\rangle =A\left\vert
A\right\rangle $, one can write%
\begin{equation*}
\left\vert \psi (t)\right\rangle =
\end{equation*}%
\begin{equation*}
\sqrt{1-\gamma ^{2}}e^{\mu _{3}+\frac{|A|^{2}}{2}\left( e^{2\mu
_{2}}-1\right) }e^{-\frac{i}{\hslash }ge^{\mu _{2}}At\hat{b}_{2}}e^{e^{\mu
_{1}-\mu _{2}}\gamma \hat{b}_{1}^{\dagger }\hat{b}_{2}^{\dagger }}\left\vert
0_{R}\right\rangle \left\vert e^{\mu _{2}}A\right\rangle .
\end{equation*}%
Using the Baker--Hausdorff formula $e^{\hat{A}}e^{\hat{B}}=e^{[\hat{A},\hat{B%
}]}e^{\hat{B}}e^{\hat{A}}$, we finally obtain%
\begin{equation*}
\left\vert \psi (t)\right\rangle =
\end{equation*}%
\begin{equation}
\sqrt{1-\gamma ^{2}}e^{\mu _{3}+\frac{|A|^{2}}{2}\left( e^{2\mu
_{2}}-1\right) }e^{-\frac{i}{\hslash }\gamma gAe^{\mu _{1}}t\hat{b}%
_{1}^{\dagger }}e^{e^{\mu _{1}-\mu _{2}}\gamma \hat{b}_{1}^{\dagger }\hat{b}%
_{2}^{\dagger }}\left\vert 0_{R}\right\rangle \left\vert e^{\mu
_{2}}A\right\rangle .  \label{v9}
\end{equation}

Using Eqs. (\ref{s3}) and (\ref{s4}), we find that the average number of
Rindler photons in the state (\ref{v9}) is 
\begin{equation*}
\left\langle \hat{b}_{1}^{\dagger }\hat{b}_{1}\right\rangle =\frac{\tilde{%
\gamma}^{2}}{1-\tilde{\gamma}^{2}}+\frac{\left( \gamma gAt\right) ^{2}}{%
\hslash ^{2}}\frac{e^{2\mu _{1}}}{\left( 1-\tilde{\gamma}^{2}\right) ^{2}},
\end{equation*}%
\begin{equation*}
\left\langle \hat{b}_{2}^{\dagger }\hat{b}_{2}\right\rangle =\frac{\tilde{%
\gamma}^{2}}{1-\tilde{\gamma}^{2}}+\frac{\left( \gamma gAt\right) ^{2}}{%
\hslash ^{2}}\frac{e^{2\mu _{1}}\tilde{\gamma}^{2}}{\left( 1-\tilde{\gamma}%
^{2}\right) ^{2}},
\end{equation*}%
where 
\begin{equation*}
\tilde{\gamma}=e^{\mu _{1}-\mu _{2}}\gamma .
\end{equation*}%
The average number of oscillator excitations in the state (\ref{v9}) is 
\begin{equation*}
\left\langle \hat{\sigma}^{\dagger }\hat{\sigma}\right\rangle =e^{2\mu
_{2}}A^{2}.
\end{equation*}

Constraints $\left\langle \hat{\sigma}^{\dagger }\hat{\sigma}-\hat{b}%
_{2}^{\dagger }\hat{b}_{2}\right\rangle =const$ and $\left\langle \hat{b}%
_{1}^{\dagger }\hat{b}_{1}\right\rangle =const$ give equations%
\begin{equation*}
e^{2\mu _{2}}A^{2}-\frac{\tilde{\gamma}^{2}}{1-\tilde{\gamma}^{2}}-\frac{%
\left( \gamma gAt\right) ^{2}}{\hslash ^{2}}\frac{e^{2\mu _{1}}\tilde{\gamma}%
^{2}}{\left( 1-\tilde{\gamma}^{2}\right) ^{2}}=A^{2}-\frac{\gamma ^{2}}{%
1-\gamma ^{2}},
\end{equation*}%
\begin{equation*}
\frac{\tilde{\gamma}^{2}}{1-\tilde{\gamma}^{2}}+\frac{\left( \gamma
gAt\right) ^{2}}{\hslash ^{2}}\frac{e^{2\mu _{1}}}{\left( 1-\tilde{\gamma}%
^{2}\right) ^{2}}=\frac{\gamma ^{2}}{1-\gamma ^{2}},
\end{equation*}%
which for $t\rightarrow \infty $ yield%
\begin{equation*}
\tilde{\gamma}\rightarrow 0,\quad \frac{1}{\hslash }\gamma gAe^{\mu
_{1}}\approx \frac{\gamma }{\sqrt{1-\gamma ^{2}}t},\quad e^{2\mu
_{2}}A^{2}\rightarrow A^{2}-\frac{\gamma ^{2}}{1-\gamma ^{2}}.
\end{equation*}%
Therefore, for $t\rightarrow \infty $%
\begin{equation*}
\left\langle \hat{b}_{1}^{\dagger }\hat{b}_{1}\right\rangle =\frac{\gamma
^{2}}{1-\gamma ^{2}},
\end{equation*}%
\begin{equation*}
\left\langle \hat{b}_{2}^{\dagger }\hat{b}_{2}\right\rangle =0,
\end{equation*}%
\begin{equation*}
\left\langle \hat{\sigma}^{\dagger }\hat{\sigma}\right\rangle =A^{2}-\frac{%
\gamma ^{2}}{1-\gamma ^{2}},
\end{equation*}%
and the normalized state vector of the system is%
\begin{equation*}
\left\vert \psi (\infty )\right\rangle =e^{-\frac{\gamma ^{2}}{2\left(
1-\gamma ^{2}\right) }}e^{-i\frac{\gamma }{\sqrt{1-\gamma ^{2}}}\hat{b}%
_{1}^{\dagger }}\left\vert 0_{R}\right\rangle \left\vert \sqrt{A^{2}-\frac{%
\gamma ^{2}}{1-\gamma ^{2}}}\right\rangle .
\end{equation*}%
The final state is the Rindler vacuum for photons $\hat{b}_{2}$, a coherent
state for photons $\hat{b}_{1}$ with the average photon number $\gamma
^{2}/(1-\gamma ^{2})$, and the oscillator remains in the coherent state with
a reduced average excitation number $A^{2}-\gamma ^{2}/(1-\gamma ^{2})$.

\end{document}